\documentclass[letterpaper]{article} % DO NOT CHANGE THIS
\usepackage{aaai25}  % DO NOT CHANGE THIS
\usepackage{times}  % DO NOT CHANGE THIS
\usepackage{helvet}  % DO NOT CHANGE THIS
\usepackage{courier}  % DO NOT CHANGE THIS
\usepackage[hyphens]{url}  % DO NOT CHANGE THIS
\usepackage{graphicx} % DO NOT CHANGE THIS
\usepackage{natbib}  % DO NOT CHANGE THIS AND DO NOT ADD ANY OPTIONS TO IT
\usepackage{caption} % DO NOT CHANGE THIS AND DO NOT ADD ANY OPTIONS TO IT
\usepackage{algorithm}
\usepackage{algorithmic}

\usepackage{newfloat}
\usepackage{listings}
\DeclareCaptionStyle{ruled}{labelfont=normalfont,labelsep=colon,strut=off} % DO NOT CHANGE THIS
\floatstyle{ruled}
\newfloat{listing}{tb}{lst}{}
\floatname{listing}{Listing}
\usepackage{amsmath}
\usepackage{booktabs}

\usepackage{tikz}
\usetikzlibrary{math, fit}

\tikzstyle{input}           = [inner sep=1pt]
\tikzstyle{gate}            = [circle, fill=white, draw=black, minimum size=4mm, inner sep=0pt]
\tikzstyle{outgate}         = [gate, thick]
\tikzstyle{wire}            = [draw,->]
\tikzstyle{notwire}         = [draw,->,dashed]

\title{Cirbo: A New Tool for Boolean Circuit Analysis and Synthesis}

\author {
	Daniil Averkov\textsuperscript{\rm 1},
	Tatiana Belova\textsuperscript{\rm 2, \rm 3},
	Gregory Emdin\textsuperscript{\rm 4},
	Mikhail Goncharov\textsuperscript{\rm 5, \rm 6},
	Viktoriia Krivogornitsyna\textsuperscript{\rm 2},
	Alexander~S. Kulikov\textsuperscript{\rm 6},
	Fedor Kurmazov\textsuperscript{\rm 2},
	Daniil Levtsov\textsuperscript{\rm 5},
	Georgie Levtsov\textsuperscript{\rm 5},
	Vsevolod Vaskin\textsuperscript{\rm 5},
	Aleksey Vorobiev\textsuperscript{\rm 3}
}
\affiliations {
	\textsuperscript{\rm 1}St.~Petersburg State University\\
	\textsuperscript{\rm 2}Steklov Mathematical Institute at St.~Petersburg, Russian Academy of Sciences\\
	\textsuperscript{\rm 3}ITMO University\\
	\textsuperscript{\rm 4}École Polytechnique Fédérale de Lausanne\\
	\textsuperscript{\rm 5}Neapolis University Pafos\\
	\textsuperscript{\rm 6}JetBrains Research\\
}

\begin{document}
\maketitle

\begin{abstract}
    We~present an~open-source tool for manipulating Boolean circuits. It~implements efficient algorithms, both existing and novel, for a~rich variety
    of~frequently used circuit tasks such~as satisfiability, synthesis, and minimization.
    We~tested the tool on a~wide range of~practically relevant
    circuits (computing, in~particular, symmetric and arithmetic functions) that have been optimized intensively by~the community
    for the last three years.
    The tool helped~us to~win the IWLS 2024 Programming Contest.
    In~2023, it~was Google DeepMind who took the first place in~the competition.
    We~were able to~reduce the size
	of~the best circuits from 2023
    by~12\% on~average, whereas for some individual circuits,
    our size reduction was as large as 83\%.
\end{abstract}

\begin{links}
    \link{Code}{https://github.com/SPbSAT/cirbo}
\end{links}

\section{Introduction}
Boolean circuits is a~mathematical model with applications
in~various branches of~Computer Science such~as Complexity Theory,
Computer Engineering, and Cryptography. The two most important related
computational problems
are circuit analysis
and circuit synthesis.
\begin{description}
	\item[Circuit analysis:] given a~circuit, check whether
	it~possesses a certain property. A~ubiquitous property is~satisfiability (whether it~is possible
	to~assign 0/1 values to~the inputs such that the circuit evaluates to~1). The corresponding problem is~known as~Circuit SAT.
	On~the one hand, many circuit analysis problems (such~as
	logical equivalence checking and verification) are equivalent
	to~Circuit SAT. On~the other hand, Circuit SAT is a~generalization
	of~SAT (satisfiability of~formulas in~conjunctive normal form)
	and many hard combinatorial problems are reduced to~SAT via Circuit SAT.
	At~the same time, Circuit SAT can be~reduced to~SAT
	in a~natural way. This way, Circuit SAT shares a~wide range
	of~applications, both practical and theoretical, with SAT~\cite{DBLP:series/faia/336}.

	\item[Circuit synthesis:] given a~specification of a~Boolean function, synthesize a~small circuit computing this function.
	This is~the first step of~integrated circuit design,
	a~highly important problem in~practice.
	In~Theoretical Computer Science, this problem is~known~as the Minimum Circuit Size Problem (MCSP): given a~truth table of a~Boolean function and
	an~integer parameter~$r$, check whether the function can
	be~computed by a~circuit of~size~$r$. MCSP is~arguably
	as~important as~SAT~\cite{DBLP:conf/fsttcs/Santhanam22}.

	Proving that a~given Boolean function cannot be~computed
	by a~small circuit (that~is, proving circuit lower bounds)
	is~notoriously hard. The number of~functions and the number
	of~circuits (as~functions of~the number of~inputs) grow too fast
	making~it infeasible in~practice to~show that a~given function
	on, say, 10~inputs cannot be~computed by~a~circuit with 40~gates.
	Moreover, currently, there are no~methods that allow
	one to~exclude a~possibility that every problem from~NP
	has circuit size at~most~$4n$, where, as~usual, $n$~denotes the input size
	\cite{DBLP:conf/focs/FindGHK16,DBLP:conf/stoc/Li022,DBLP:journals/cc/FindGHK23}.
\end{description}

\subsection{New Tool}
The focus of~the this paper
is~on~practical aspects of~the two problems mentioned above: we~present a~new tool,
called \texttt{Cirbo},
for solving a~wide range of~problems on~Boolean circuits.
The tool implements a~variety
of~algorithms, both known and novel ones.
The tool allowed~us to~win
the IWLS 2024 Programming Contest.\footnote{\url{https://www.iwls.org/iwls2024/}}
The goal of~the competition is~to~synthesize efficient circuits
for $100$~Boolean functions (specified by~their truth tables),
in~two bases, XAIG and AIG. For each of~the two bases,
for more than half of~the functions, the circuits synthesized
by~our tool turned out to~be the smallest among the circuits produced by~all teams. Moreover, the datasets in~2024 contest were the same
as~in 2023, giving~us a~possibility to~track the progress
of~reducing the circuit size for these datasets. Table~\ref{table:datasets} shows the corresponding circuit size for a~selection of~datasets and highlights that
in~some cases our size reduction was as~large as~83\%.
Later in~the
text, we~define all the functions from the table formally,
provide more statistics as~well~as detailed steps that led
to~improved circuits.

\begin{table*}[t]
	\caption{Circuit size for a~selection of~benchmarks from IWLS 2024 Programming Contest from various categories.
	For each benchmark, we~show, in~the first two columns, its short description
	as~well~as the name of~the corresponding benchmark.
	The next two columns
	show the smallest circuit size in~the AIG basis found in~2023 as~well as~the size of a~circuit
	synthesized by~our tool in~2024.
	The next column shows size reduction
	in~percent (compared to~the best circuit size in~2023).
	Finally, the last three columns show the same data for the XAIG basis.}
	\label{table:datasets}
	\begin{center}
	\begin{tabular}{lcccccccccc}
		\toprule
		& & \multicolumn{3}{c}{AIG} & \multicolumn{3}{c}{XAIG}\\
		\cmidrule(lr){3-5}  \cmidrule(lr){6-8}
		function & IWLS 2024 code & 2023 (best) & 2024 (our) & improvement & 2023 (best) & 2024 (our) & improvement\\
		\midrule
		modulo8 & ex33 & 1182 & 250 & 78.85\% & 1158 & 190 & 83.59\%\\
		square12 & ex96  & 1319 & 476 & 63.91\% & 1284 & 324 & 74.77\%\\
		div8 & ex88  & 317 & 176 & 44.48\% & 306 & 142 & 53.59\%\\
		sqrt16 & ex65  & 239 & 176 & 26.36\% & 226 & 136 & 39.82\%\\
		sort15 & ex39  & 114 & 96 & 15.79\% & 114 & 73 & 35.96\%\\
		maj15 & ex36  & 68 & 68 & 0\% & 68 & 48 & 29.41\%\\
		neuron & ex20  & 653 & 588 & 9.95\% & 644 & 587 & 8.85\%\\
		espresso & ex25  & 1381 & 1340 & 2.97\% & 1381 & 1339 & 3.04\%\\
		\bottomrule
	\end{tabular}
	\end{center}
\end{table*}

\subsection{Related Work}
There is a~number of~packages providing a~similar functionality:
\texttt{ABC}~\cite{DBLP:conf/cav/BraytonM10} and \texttt{mockturtle}~\cite{DBLP:journals/corr/abs-1805-05121}
are general purpose tools for working with Boolean circuits implemented in~\texttt{C++}, whereas \texttt{CLI}~\cite{DBLP:conf/mfcs/KulikovPS22}
and \texttt{CIOPS}~\cite{DBLP:conf/aaai/ReichlSS23} are circuit minimization tools based on~SAT/QBF solvers implemented in~\texttt{Python}. As~our experiments show, our tool
is~capable of~solving various datasets better than the tools mentioned
above. At~the same time, for some of~these datasets,
the best results have been achieved by~combining our tool with the existing ones.

\section{General Setting}
%\subsection{Notation}
For a~predicate~$P$, $[P]$ is~the Iverson bracket: $[P]=1$ if $P$~is true and $[P]=0$ otherwise. For a~non-negative integer~$q$,
$\operatorname{bin}(q)$ is~the binary representation of~$q$
(padded with a~number of~leading zeroes if~needed).
Conversely, for a~bit-string $b=(b_0,\dotsc, b_k)$, $\operatorname{int}(b)=\sum_{i=0}^{k}2^ib_i$ is~the corresponding integer.

\subsection{Boolean Functions}
Let $B_{n,m}=\{f \colon \{0,1\}^n \to \{0,1\}^m\}$ be~the set of~all Boolean functions with $n$~inputs and $m$~outputs
and let $B_n=B_{n,1}$ be the set of~all $n$-input single-output functions
(that is, predicates).
A~function of the form $f \colon \{0,1\}^n \to \{0,1,*\}^m$
is~called \emph{partially defined}: $*$ is~known~as \emph{don't care} symbol
and means an~undefined Boolean value.

Below, we~define a~number of~specific Boolean functions studied in~this paper.
By $x=(x_1,\dotsc, x_n)$~and~$y=(y_1, \dotsc, y_n)$ we~denote input $n$-bit strings
and $\operatorname{sum}(x)=x_1+\dotsb+x_n$.
\begin{itemize}
	\item $\operatorname{MAJ}_n \in B_n$ is~the majority function, that is, it~is equal to~$1$ if~and only~if more than half of~the $n$~input bits are~$1$'s:
	\(\operatorname{MAJ}_n(x)=[\operatorname{sum}(x) > n/2].\)
	\item $\operatorname{SUM}_n \in B_{n, \lceil \log_2(n+1) \rceil}$ computes the binary representation of~the sum of~$n$~input bits:
	\(\operatorname{SUM}_n(x)=\operatorname{bin}(\operatorname{sum}(x)).\)
	\item $\operatorname{SORT}_n \in B_{n,n}$ sorts the given $n$~bits:
	\(\operatorname{SORT}_n(x)=(x_1', \dotsc, x_n'),\)
	where $x_1' \le \dotsb \le x_n'$ and $\operatorname{sum}(x)=\operatorname{sum}(x')$.
	\item $\operatorname{MULT}_n \in B_{2n, 2n}$ computes the product of~the given two~$n$-bit integers:
	\(\operatorname{MULT}_n(x, y)=\operatorname{bin}(\operatorname{int}(x) \cdot \operatorname{int}(y)).\)
	\item $\operatorname{SQR}_n \in B_{n, 2n}$ computes the square of~the given $n$-bit integer:
	\(\operatorname{SQR}_n(x)=\operatorname{MULT}_n(x, x).\)
	\item $\operatorname{SQRT}_n \in B_{n, n/2}$ computes the square root of~the given $n$-bit integer:
	\(\operatorname{SQRT}_n(x)=\operatorname{bin}(\lfloor \sqrt{\operatorname{int}(x)}\rfloor).\)
	\item $\operatorname{DIV}_n \in B_{2n, n}$ and $\operatorname{MOD}_n \in B_{2n, n}$
	functions compute, respectively, the quotient and the remainder of~the first input integer divided by~the second input integer:
	\begin{align*}
		\operatorname{DIV}_n(x,y)&=\operatorname{bin}(\lfloor \operatorname{int}(x) /\operatorname{int(y)}\rfloor),\\
		\operatorname{MOD}_n(x,y)&=\operatorname{bin}(\operatorname{int}(x) \bmod \operatorname{int(y)}).
	\end{align*}
\end{itemize}

\subsection{Boolean Circuits}
A~circuit is~a~natural way of~computing Boolean functions.
It~is an~acyclic directed graph of in-degree at~most~$2$ whose $n$~source
nodes are labeled with input variables
$x_1, \dotsc, x_n$ and all other nodes (called \emph{internal})
are labeled with Boolean operations from
$B_1 \cup B_2$ (that~is, unary and binary Boolean predicates).
The nodes of~the circuit are called \emph{gates} and each gate computes
a~(single-output) Boolean function of~$x_1, \dotsc, x_n$. Thus, if~$m$ gates of the
circuit are marked as~outputs, it~computes a~function from $B_{n,m}$.
The size of a~circuit is~its number of~internal binary gates
(it~is common to~assume that unary gates are given for free).

\begin{figure}[tb]
	\caption{A~circuit (known as~Full Adder) and the corresponding straight line program
		(in~\texttt{Python}) for $\operatorname{SUM}_3$. The output gates are shown in~bold.}
	\label{fig:fulladder}
	\begin{center}
		\begin{tikzpicture}[label distance=-.9mm, yscale=.7]
			%\draw[help lines] (0, 0) grid (6, 6);
			\foreach \n/\x/\y in {1/0/3, 2/1/3, 3/2/3}
				\node[input] (x\n) at (\x, \y) {$x_{\n}$};
			\node[gate, label=left:$a$] (g1) at (0.5,2) {$\oplus$};
			\node[gate, label=left:$b$] (g2) at (1.5,2) {$\oplus$};
			\node[gate, label=left:$c$] (g3) at (0.5,1) {$\lor$};
			\node[outgate, label=right:$w_0$] (g4) at (1.5,1) {$\oplus$};
			\node[outgate, label=right:$w_1$] (g5) at (0.5,0) {$\oplus$};
			\foreach \f/\t in {x1/g1, x2/g1, x2/g2, x3/g2, g1/g3, g2/g3, g1/g4, g3/g5, g4/g5}
				\draw[wire] (\f) -- (\t);
			\path (x3) edge[bend left, wire] (g4);

			\node (c) at (0.5, -1) {\strut carry};
			\node (s) at (1.5, -1) {\strut sum};
			\draw[wire] (g4) -- (s);
			\draw[wire] (g5) -- (c);

			\node[right] at (3, 1) {
				\begin{lstlisting}[language=Python]
def sum3(x1, x2, x3):
	a = x1 ^ x2
	b = x2 ^ x3
	c = a | b
	w0 = a ^ x3
	w1 = c ^ w0
	return w0, w1
				\end{lstlisting}
			};
		\end{tikzpicture}
	\end{center}
\end{figure}
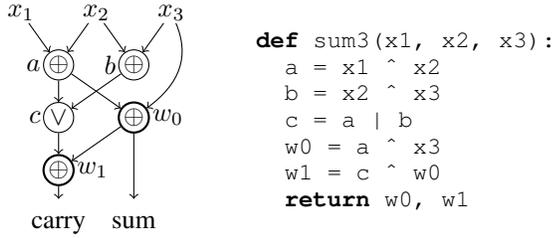

Figure~\ref{fig:fulladder} shows an~example of~a~circuit of~size~$5$ computing
$\operatorname{SUM}_3$. It~also highlights that a~circuit corresponds
to~an~extremely simple program (called a~straight line program):
every line of~this program just applies a~unary or~binary Boolean operation
to~input bits or~the results of~the previous lines.

We~assume that a~gate of~a~circuit can compute any unary or~binary Boolean function.
It~is not difficult to~see that, for a~given Boolean function~$f$,
the minimum size of a~circuit computing~$f$ is~equal
to~the minimum size of a~circuit for~$f$
when each gate computes either
binary XOR ($\oplus$), binary AND ($\land$), or unary NOT ($\neg$):
indeed, every binary Boolean operation is~either a~summation
or a~multiplication with possibly negated inputs and outputs. For this reason,
when all unary and binary operations are allowed, we~say that this is an~XAIG circuit:
X~stands for XOR (summation),
A~stands for AND (multiplication),
I~stands for inverter (negation),
and G~stands for a~graph.

\begin{figure}[tb]
	\caption{A~circuit over the basis $B_2 \setminus \{\oplus, \equiv\}$ computing
	$\operatorname{SUM}_3$ (left) and its AIG representation (right). The output gates are shown in~bold, whereas the negated wires are~shown dashed. The binary Boolean operation~$>$
	is~defined in a~natural way: $a>b=a \land \overline{b}$.}
	\label{fig:fulladderaig}
	\begin{center}
		\begin{tikzpicture}[yscale=.7]
			%\draw[help lines] (0, 0) grid (6, 6);
			\foreach \n/\x/\y in {1/0/1, 2/1/3, 3/2/3}
				\node[input] (x\n) at (\x, \y) {$x_{\n}$};
			\foreach \n/\x/\y/\op/\p/\q in {1/2/2/\land/x2/x3, 2/1/2/\lor/x2/x3, 3/1/1/>/1/2, 4/0/0/\lor/x1/3, 5/1/0/\land/x1/3} {
				\node[gate] (\n) at (\x, \y) {$\op$};
				\draw[wire] (\p) -- (\n);
				\draw[wire] (\q) -- (\n);
			}
			\foreach \n/\x/\y/\op/\p/\q in {6/0/-1/>/4/5, 7/2/-1/\lor/1/5} {
				\node[outgate] (\n) at (\x, \y) {$\op$};
				\draw[wire] (\p) -- (\n);
				\draw[wire] (\q) -- (\n);
			}
			%\node at (1, -2.5) {(a)};

			\node (c) at (2, -2) {\strut carry};
			\node (s) at (0, -2) {\strut sum};
			\draw[wire] (6) -- (s);
			\draw[wire] (7) -- (c);

			\begin{scope}[xshift=40mm]
				\foreach \n/\x/\y in {1/0/1, 2/1/3, 3/2/3}
					\node[input] (x\n) at (\x, \y) {$x_{\n}$};
				\foreach \n/\x/\y/\op/\p/\wp/\q/\wq in {
					1/2/2/\land/x2/wire/x3/wire,
					2/1/2/\lor/x2/notwire/x3/notwire,
					3/1/1/>/1/notwire/2/notwire,
					4/0/0/\lor/x1/notwire/3/notwire,
					5/1/0/\land/x1/wire/3/wire}
				{
					\node[gate] (\n) at (\x, \y) {};
					\draw[\wp] (\p) -- (\n);
					\draw[\wq] (\q) -- (\n);
				}
				\foreach \n/\x/\y/\op/\p/\wp/\q/\wq in {
					6/0/-1/>/4/notwire/5/notwire,
					7/2/-1/\lor/1/notwire/5/notwire}
				{
					\node[outgate] (\n) at (\x, \y) {};
					\draw[\wp] (\p) -- (\n);
					\draw[\wq] (\q) -- (\n);
				}
			%\node at (1, -2.5) {(b)};

			\node (c) at (2, -2) {\strut carry};
			\node (s) at (0, -2) {\strut sum};
			\draw[wire] (6) -- (s);
			\draw[notwire] (7) -- (c);
			\end{scope}
		\end{tikzpicture}
	\end{center}
\end{figure}
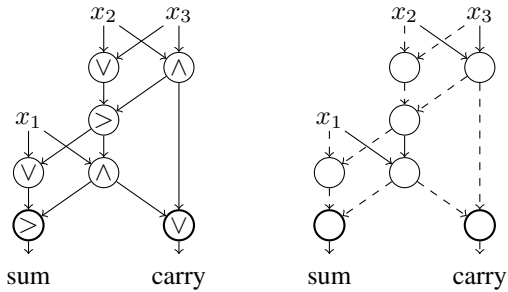

It~is~well known that any Boolean function can also be~computed
by~a~circuit that only uses AND's and NOT's as~operations in~gates.
They are called AIG circuits \cite{BiereHeljankoWieringa2011} and this is a~convenient format for representing
a~circuit in~practice: since every binary gate computes an~AND, one just stores
a~graph of~in-degree~$2$ (without storing the operations computed in~the gates)
and a~list of~flags telling which of~the edges are negated (or~inverted).
See an~example in~Figure~\ref{fig:fulladderaig}.

\section{Tool Features}
The tool has been implemented with a~goal
of~being
efficient and easy~to~use and extend: the code is~open source\footnote{The code is publicly available at \url{https://github.com/SPbSAT/cirbo}.}
and written in~\texttt{Python} (making the code compact and easy to~read).
Below, we~describe the main features of~the tool.
For many of~them, we~complement their description
with a~short code snippet to~highlight the ease of~use.
In~the code samples below, we~omit the preamble
(that loads the necessary packages) to~save space.
The complete code snippets (runnable out of~the box)
can be~found in~the \texttt{tutorial} folder
of~the supplementary archive file.

\subsection{Analysis}
The tool allows to~analyze Boolean functions and circuits.

\subsubsection{Function Analysis}
Analyzing the properties of a~given Boolean function
is~important for subsequent synthesis of an~efficient
circuit computing this function. Currently, the tool allows
to~check whether a~Boolean function is~monotone or~symmetric
(these checks are performed via enumerating all $2^n$
input assignments and hence are only practical when the number~$n$
of~inputs is~small enough). A~function can be~passed either as~a~truth table
or as a~\texttt{Python} function, see Listing~\ref{lst:analysis}.

\begin{listing}[tb]
    \caption{Analyzing Boolean functions.}
    \label{lst:analysis}
\begin{lstlisting}[language=Python, linerange={4-12}]
f = PyFunction(lambda xs: [sum(xs) % 2], 4)
print(f.is_monotone())

g = PyFunction.from_int_binary_func(
    lambda x, y: x + y, 2, 3)
print(g.is_symmetric())

e = TruthTable(['1101'])
print(e.is_constant())
\end{lstlisting}
\end{listing}

\subsubsection{Circuit Analysis}
A~circuit is a~particular way of~representing a~Boolean function.
For this reason, the \texttt{Circuit} class implements \texttt{Function} interface
allowing one to~use all the checks described in the previous section
for circuits also. Additionally, one can check whether a~circuit is
satisfiable. This is~done by~transforming a~circuit into a~CNF (via Tseitin transformation~\cite{Tseitin70})
and invoking a~SAT solver via the \texttt{pysat} module~\cite{DBLP:conf/sat/IgnatievMM18},
see Listing~\ref{lst:sat}.
\begin{listing}[tb]
    \caption{Checking whether a~circuit is~satisfiable.}
    \label{lst:sat}
\begin{lstlisting}[language=Python, linerange={5-8}]
path = '../data/circuit.bench'
ckt = Circuit.from_bench_file(path)
result = is_circuit_satisfiable(ckt)
print(result.answer)
\end{lstlisting}
\end{listing}

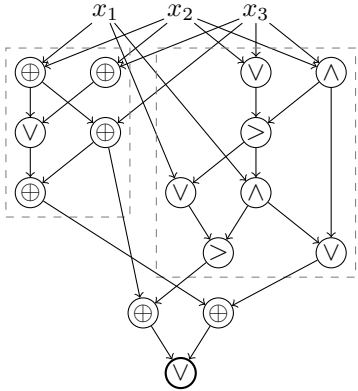
\begin{figure}
	\caption{A~miter composed out of~circuits from Figures~\ref{fig:fulladder} and~\ref{fig:fulladderaig}. Since these two circuits compute the same function,
		the miter is~unsatisfiable.}
	\label{fig:miter}
	\begin{center}
		\begin{tikzpicture}[label distance=-.9mm, yscale=.8]
			\foreach \n/\x/\y in {1/1.5/3, 2/2.5/3, 3/3.5/3}
			\node[input] (x\n) at (\x, \y) {$x_{\n}$};

			\node[gate] (g1) at (0.5,2) {$\oplus$};
			\node[gate] (g2) at (1.5,2) {$\oplus$};
			\node[gate] (g3) at (0.5,1) {$\lor$};
			\node[gate] (g4) at (1.5,1) {$\oplus$};
			\node[gate] (g5) at (0.5,0) {$\oplus$};
			\foreach \f/\t in {x1/g1, x2/g1, x2/g2, x3/g2, g1/g3, g2/g3, g1/g4, g3/g5, g4/g5}
			\draw[wire] (\f) -- (\t);
			\path (x3) edge[wire] (g4);

			\begin{scope}[xshift=25mm]
				\foreach \n/\x/\y/\op/\p/\q in {1/2/2/\land/x2/x3, 2/1/2/\lor/x2/x3, 3/1/1/>/1/2, 4/0/0/\lor/x1/3, 5/1/0/\land/x1/3} {
					\node[gate] (\n) at (\x, \y) {$\op$};
					\draw[wire] (\p) -- (\n);
					\draw[wire] (\q) -- (\n);
				}
				\foreach \n/\x/\y/\op/\p/\q in {6/0.5/-1/>/4/5, 7/2/-1/\lor/1/5} {
					\node[gate] (\n) at (\x, \y) {$\op$};
					\draw[wire] (\p) -- (\n);
					\draw[wire] (\q) -- (\n);
				}
			\end{scope}

            \node[draw=gray, dashed, fit=(g1) (g2) (g3) (g4) (g5)] {};
            \node[draw=gray, dashed, fit=(1) (2) (3) (4) (5) (6) (7)] {};

			\node[gate] (m1) at (2, -2) {$\oplus$};
			\node[gate] (m2) at (3, -2) {$\oplus$};
			\node[outgate] (m3) at (2.5, -3) {$\lor$};
			\foreach \f/\t in {m1/m3, m2/m3, g4/m1, 6/m1, g5/m2, 7/m2}
			\draw[wire] (\f) -- (\t);
		\end{tikzpicture}
	\end{center}
\end{figure}

\begin{figure}[tb]
	\caption{A~visualization of the miter circuit produced by~Listing~\ref{lst:verification}.}
	\label{fig:automiter}
	\begin{center}
		\includegraphics[width=.6\linewidth]{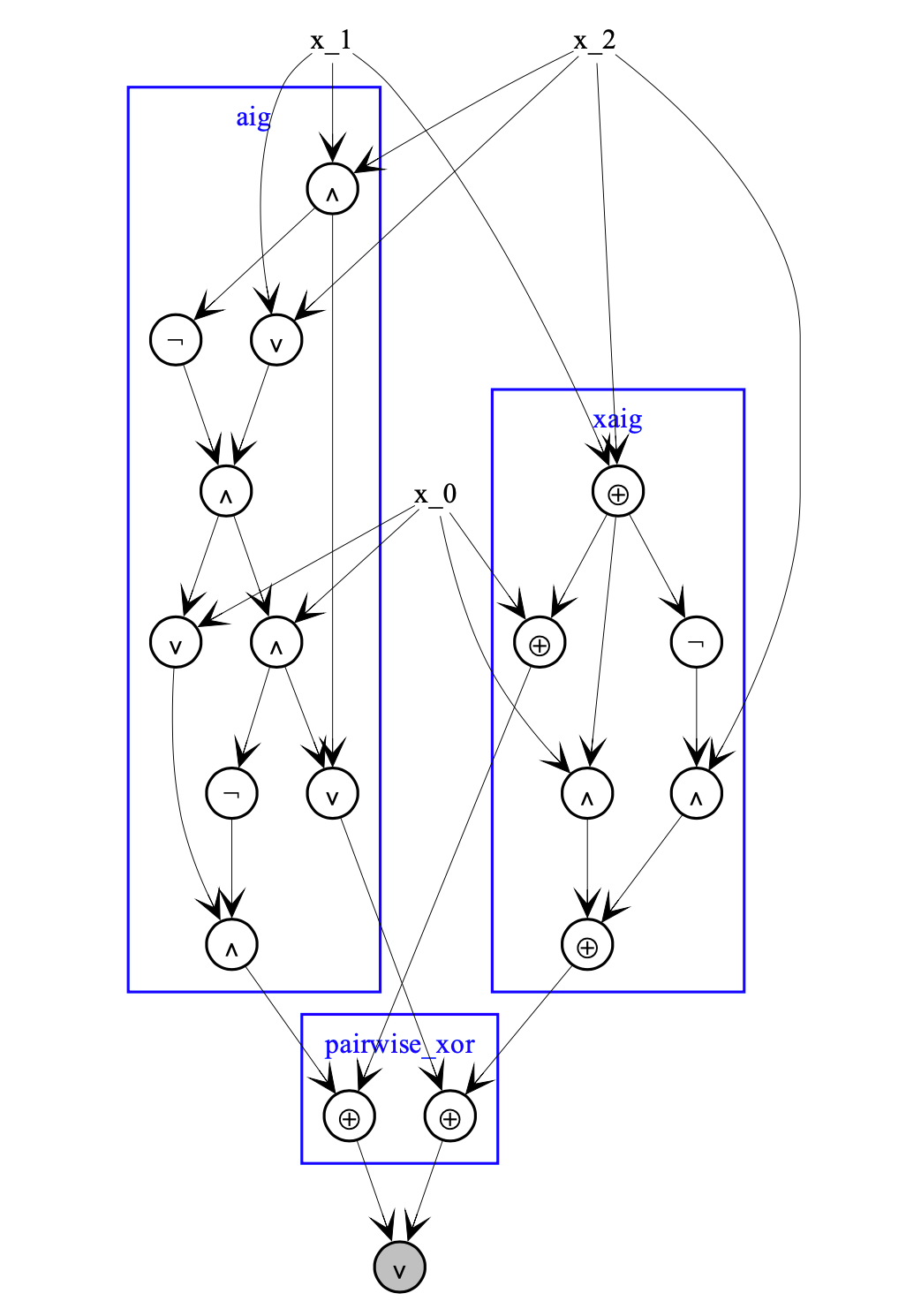}
	\end{center}
\end{figure}

Circuit satisfiability is a~ubiquitous problem as~many other hard problems
can be~reduced to~it naturally. For example, to~verify whether two
circuits $C_1, C_2 \colon \{0,1\}^n \to \{0,1\}^m$ compute the same
function, one combines them into a~miter circuit
\[M(x)=\bigvee_{i=1}^{m}(y_i \oplus z_i)\]
where $(y_1, \dots, y_m)=C_1(x)$ and $(z_1, \dots, z_m)=C_2(x)$).
One then
checks whether $M$~is~satisfiable (it~is if~and only~if $C_1$~and~$C_2$ do~not
compute the same function).
Figure~\ref{fig:miter} gives an~example.
A~miter can also be~constructed easily
in~the tool, see Listing~\ref{lst:verification}.
\begin{listing}[tb]
	\caption{Verifying that two circuits compute the same function.}
	\label{lst:verification}
\begin{lstlisting}[language=Python, linerange={6-20}]
aig = generate_sum_n_bits(3, basis='AIG')
xaig = generate_sum_n_bits(3, basis='XAIG')

mtr = Circuit().add_circuit(aig, name='aig')
mtr.connect_inputs(xaig, name='xaig')
mtr.extend_circuit(
    generate_pairwise_xor(aig.output_size),
    name='pairwise_xor',
)  

outs = mtr.get_block('pairwise_xor').outputs      
mtr.emplace_gate('or', OR, tuple(outs))
mtr.set_outputs(['or'])

mtr.view_graph(autorename_labels=True)
\end{lstlisting}
\end{listing}
The listing also illustrates two other important features of~the tool.
First, one can compose a~circuit out of~blocks. While designing a~circuit, reasoning in~terms of~blocks, rather than in~terms of~gates, is~more convenient. Second, one can draw a~circuit. The drawing method also shows the block structure of~the circuit.
Figure~\ref{fig:automiter} shows the resulting drawing of~the miter circuit.

One can also reduce
many other hard combinatorial problems to~circuit satisfiability. We~illustrate
this for the
factorization problem where one is~given a~positive integer~$k$
and is~asked to~find its nontrivial divisor (or~to~report that $k$~is~prime).
Listing~\ref{lst:factorization} shows such a~reduction. Using
generator for multiplying and equality,
the reduction constructs a~circuit that is~satisfiable if~and only~if
there exists $(n-1)$-bit non-negative integers $p$~and~$q$ (where $n$~is~the bit length of~$k$) such that $pq=k$.
If~the resulting circuit is~satisfiable, the values of~$p$ and~$q$ can be~read off
from its satisfying assignment.

\begin{listing}[tb]
    \caption{Reducing the factorization problem to~circuit satisfiability.}
    \label{lst:factorization}
\begin{lstlisting}[language=Python, linerange={7-15}]
def factorization(number: int) -> Circuit:
    n = number.bit_length()
    ckt = Circuit.bare_circuit(input_size=2 * (n - 1))
    p, q = ckt.inputs[:n - 1], ckt.inputs[n - 1:]
    outs = add_mul(ckt, q, p)
    g3 = add_equal(ckt, outs, number)
    ckt.mark_as_output(g3)
    return ckt
\end{lstlisting}
\end{listing}

\subsection{Synthesis}
Using the tool, one can synthesize Boolean circuits using the following three regimes.

\subsubsection{Manual Synthesis out~of~Presynthesized Blocks}
\label{sec:manual}
The tool contains
generators of~various circuits that are frequently used
in~circuit synthesis: comparators, summators, multipliers, etc.
One can use them as~building blocks to~synthesize efficient circuits for various functions.
We~give an~example for the majority function of~six input bits.
This is a~symmetric function, so one can first compute the binary representation
$(b_0, b_1, b_2)$ of~the sum of~the input bits and then output $(b_0 \land b_1) \lor b_2$.
Listing~\ref{lst:majority} shows how to~achieve this in~the tool.

\begin{listing}[tb]
    \caption{Synthesizing a~circuit for the majority function.}
    \label{lst:majority}
\begin{lstlisting}[language=Python, linerange={6-11}]
ckt = Circuit.bare_circuit(input_size=6)
b0, b1, b2 = add_sum_n_bits(ckt, ckt.inputs)
ckt.add_gate(Gate('a0', AND, (b0, b1)))
ckt.add_gate(Gate('a1', OR, ('a0', b2)))
ckt.mark_as_output('a1')
ckt.view_graph()
\end{lstlisting}
\end{listing}

\subsubsection{Automated SAT-based Synthesis}
When the number~$n$ of~inputs is small (say, $n \le 10$), one can
synthesize an~efficient circuit for a~given function using a~SAT-based approach
\cite{DBLP:conf/sat/KojevnikovKY09}: one transforms the~statement
``there exists a~circuit of~the given size computing the given function'' to~CNF
and checks its validity using a~SAT solver.
Listing~\ref{lst:satbasedsynthesis} shows how one can automatically
synthesize the Full Adder circuit (shown in~Figure~\ref{fig:fulladder})
using this approach.

\begin{listing}[tb]
    \caption{Synthesizing Full Adder using SAT-based approach.}
    \label{lst:satbasedsynthesis}
\begin{lstlisting}[language=Python, linerange={5-12}]
def sum_3(x1, x2, x3):
    s = x1 + x2 + x3
    return [(s >> i) & 1 for i in range(2)]

func = PyFunction.from_positional(sum_3)
cf = CircuitFinderSat(func, 5, basis='XAIG')
ckt = cf.find_circuit()
ckt.view_graph()
\end{lstlisting}
\end{listing}

\subsubsection{Hybrid Synthesis}
One can combine the two strategies above by~synthesizing a~circuit out
of~built-in blocks and blocks synthesized via SAT-based approach.
For example, instead of~manually constructing the final block
$(b_0 \land b_1) \lor b_2$ for $\operatorname{MAJ}_6$ in~the example above,
one can synthesize~it with
the SAT-based approach, see Listing~\ref{lst:hybrid}.
The listing illustrates another important feature of the SAT-based synthesis:
it~allows to~synthesize partially defined functions:
in~this particular case, the value of~the final block on~the input $(b_0=1, b_1=1, b_2=1)$
may be~arbitrary as~the sum~of the input six bits can never be~equal to~seven.

\begin{listing}[tb]
    \caption{Synthesizing a~circuit for $\operatorname{MAJ}_6$ using a~hybrid approach.}
    \label{lst:hybrid}
\begin{lstlisting}[language=Python,  linerange={6-15}]
def geq3(x: bool, y: bool, z: bool):
    s = x + 2 * y + 4 * z
    return [DontCare] if s > 6 else [True] if s >= 3 else [False]

ckt = generate_sum_n_bits(n=6)
pfm = PyFunctionModel.from_positional(geq3)
cfs = CircuitFinderSat(pfm, 2, basis='XAIG')
geq3_ckt = cfs.find_circuit()
ckt.extend_circuit(geq3_ckt, name='geq3')
ckt.view_graph()
\end{lstlisting}
\end{listing}

\subsection{Minimization}
In the circuit minimization problem, one is~given a~circuit and
is~asked to~come~up with a~smaller circuit computing the same function.

\subsubsection{Low Effort Minimization}
The \texttt{cleanup} method performs straightforward cleaning
of~a~circuit: for example, removes dangling and duplicate gates.
To~give an~example, consider a~task of~synthesizing a~circuit for $\operatorname{MAJ}_7$.
Clearly, this is the same as~computing the most significant bit of the sum of~the input bits. Hence, it~makes sense to~compute (the binary representation~of) the sum and to~output
this bit. However, since we~only need one bit (out~of all bits of~the sum), some computations can be~dropped. This is~indeed what happens in~practice:
as~the result of~invoking the code from Listing~\ref{lst:loweffort},
two unnecessary gates
are removed from the circuit.
\begin{listing}[tb]
    \caption{Cleaning an~XAIG circuit for $\operatorname{MAJ}_7$.}
    \label{lst:loweffort}
\begin{lstlisting}[language=Python, linerange={6-11}]
ckt = Circuit.bare_circuit(input_size=7)
*_, b2 = add_sum_n_bits(ckt, ckt.inputs)
ckt.mark_as_output(b2)
print(ckt.gates_number())
ckt = cleanup(ckt)
print(ckt.gates_number())
\end{lstlisting}
\end{listing}
Also, one can run various \texttt{ABC}~commands right from the tool.
Listing~\ref{lst:abc} shows how to~achieve the same task (removing redundant gates from a~circuit for $\operatorname{SUM}_7$
to~get a~circuit for $\operatorname{MAJ}_7$) for the basis AIG
using \texttt{ABC} cleaning.
\begin{listing}[tb]
	\caption{Cleaning an~AIG circuit for $\operatorname{MAJ}_7$ using \texttt{ABC}.}
	\label{lst:abc}
\begin{lstlisting}[language=Python, linerange={5-8}]
ckt = Circuit.bare_circuit(input_size=7)
*_, lst = add_sum_n_bits(ckt, ckt.inputs, basis='AIG')
ckt.mark_as_output(lst)
ckt = abc_transform(ckt, 'strash; dc2')
\end{lstlisting}
\end{listing}

\subsubsection{High Effort Minimization}
A~more powerful, but significantly less efficient minimizing method is~the following:
try to~minimize small subcircuits of a~given circuit \cite{DBLP:conf/mfcs/KulikovPS22}.
For each subcircuit,
we~compute a~partial function computed~by~it and try to~synthesize
a~more efficient circuit computing the same function using the SAT-based approach.
If~a~more efficient circuit is~found,
we~replace the corresponding subcircuit in~the original subcircuit
and iterate.
For example, the code in~Listing~\ref{lst:higheffort}
synthesizes a~circuit of~size~$12$ out of~two Full Adders and one Half Adder.
The subsequent call
to~\texttt{minimize\_subcircuits} improves the resulting circuit by~one gate.

\begin{listing}[tb]
    \caption{SAT-based minimization of a~circuit for $\operatorname{SUM}_5$.}
    \label{lst:higheffort}
\begin{lstlisting}[language=Python, linerange={6-14}]
ckt = Circuit.bare_circuit(input_size=5)
x1, x2, x3, x4, x5 = ckt.inputs
a0, a1 = add_sum3(ckt, [x1, x2, x3])
b0, b1 = add_sum3(ckt, [a0, x4, x5])
w1, w2 = add_sum2(ckt, [a1, b1])
ckt.set_outputs([b0, w1, w2])
print(ckt.gates_number())
ckt = minimize_subcircuits(ckt, 'XAIG')
print(ckt.gates_number())
\end{lstlisting}
\end{listing}

\subsection{Database of (Nearly) Optimal Circuits}
In~the tool, we~employ a~database of~circuits
of~all Boolean functions with at~most three inputs and at~most three outputs.
The vast majority of~them are provably optimal.
Thus, each time when we~need to~synthesize an~efficient circuit,
we~first check whether it~is stored in~the database.
The number of functions~is
\[\sum_{n=2}^3 \sum_{m = 1}^3 \binom{2^{2^n}}{m}=2\,797\,112.\]
To~reduce the search space while populating the database,
we used a classification approach similar to NPN-classification
\cite{DBLP:conf/ismvl/HaaswijkTSM17}. As in NPN-classification,
we say that two functions are equivalent if one can be transformed
into the other by permuting and negating some of the inputs and outputs.
Additionally, our classification considers the permutation of outputs,
as we are working with multiple outputs.
Clearly, two equivalent functions have the same circuit size. This allows~us
to~partition the set of~all functions into equivalent classes and to~synthesize
an~efficient circuit for a~single representative from each class.

In~the XAIG basis, each function from $B_{3,3}$ has a~relatively small circuit size (at~most~$8$) and it~is possible to~find a~provably optimal circuit
via a~reduction to~SAT. We~used a~tool~\cite{DBLP:conf/mfcs/KulikovPS22} for this task.
In~the AIG basis, circuits for functions from $B_{3,3}$
can have size as~large as~$11$. For some of~these functions,
proving that there~is no~smaller circuit is~already a~difficult task
for the state-of-the-art SAT solvers.
For this reason, for several classes of~functions
we~have an~efficient circuit, but no~proof that this circuit has the smallest size.
We~provide a~detailed statistics in~Tables~\ref{table:stat}.

\begin{table}[tb]
    \centering
    \begin{tabular}{rrrrr}
        \toprule
        & \multicolumn{2}{c}{XAIG} & \multicolumn{2}{c}{AIG}\\
        \cmidrule(lr){2-3}  \cmidrule(lr){4-5}
        size & classes & functions & classes & functions \\
        \midrule
        1 & 8 & 60 & 4 & 48 \\
        2 & 74 & 2,160 & 42 & 1,461 \\
        3 & 324 & 36,672 & 142 & 17,720 \\
        4 & 1,153 & 266,500 & 373 & 81,996 \\
        5 & 2,967 & 892,312 & 949 & 241,428 \\
        6 & 3,690 & 1,242,704 & 1,759 & 515,611 \\
        7 & 1,030 & 354,528 & 2,462 & 773,088 \\
        8 & 9 & 2,176 & 2,207 & 730,576 \\
        $\le 9$ & & & 1,087 & 364,400 \\
        $\le 10$ & & & 224 & 69,664 \\
        $\le 11$ & & & 6 & 1,120 \\
        \bottomrule
    \end{tabular}
    \caption{Distribution of~classes and functions by~circuit size, for all functions with three inputs and three distinct outputs, for XAIG and AIG bases.}
    \label{table:stat}
\end{table}

\section{Experimental Evaluation}
\subsection{Synthesizing Efficient Circuits}% for Symmetric and Arithmetic Functions}
Recall that a~Boolean function $f(x_1, \dotsc, x_n) \in B_{n,m}$
is~called \emph{symmetric} if~its value depends on $(x_1+\dotsb+x_n)$
only (equivalently, the function value never changes when one permutes
the input). On~the one hand, many interesting functions are symmetric
(for example, $\operatorname{SUM}$, $\operatorname{MAJ}$, and $\operatorname{SORT}$).
On~the other hand, even if a~function is~not symmetric, a~circuit for~it
can rely on~symmetric functions: for example, to~compute the product
of~two $n$-bit integers, one first computes pairwise products of~input bits
and then computes the sums of~these products.

\subsubsection{SUM}
Since the value of~a~symmetric function depends on~$(x_1+ \dotsb + x_n)$,
$\operatorname{SUM}$ is a~fundamental symmetric function:
below, we~demonstrate that to~get an~efficient circuit computing
a~symmetric function, it~makes sense to~first compute
$(b_0, \dotsc, b_k)=\operatorname{SUM}_n(x_1, \dotsc, x_n)$ (that~is,
to~compress $n$~input bits into about $k+1=\lceil \log_2 (n+1) \rceil$ bits)
and then to~compute the result out of~$(b_0, \dotsc, b_k)$.
For this reason, it~is important to~have efficient circuits
for $\operatorname{SUM}$. A~well known way to~compute
$\operatorname{SUM}_n$ is~to~apply blocks computing $\operatorname{SUM}_3$ and $\operatorname{SUM_2}$
iteratively: in~particular, this leads to~upper bounds
\[\operatorname{size}_{\text{XAIG}}(\operatorname{SUM}_n) \le 5n \text{ and } \operatorname{size}_{\text{AIG}}(\operatorname{SUM}_n) \le 7n \, .\]
No~better construction is~known for AIG, whereas for XAIG,
a~better upper bound is~known \cite{DBLP:journals/ipl/DemenkovKKY10}: \[\operatorname{size}_{\text{XAIG}}(\operatorname{SUM}_n) \le 4.5n+o(n)\,.\]
Our tool allows to~generate best known circuits computing
$\operatorname{SUM}_n$ for all~$n$.
In~Table~\ref{table:sumn}, we~show the size of~the corresponding circuits.

\begin{table}[tb]
    \caption{Size of~currently best known circuits computing $\operatorname{SUM}_n$.}
    \label{table:sumn}
    \begin{center}
        \begin{tabular}{lrrrrrrrrrrrrrrrrrrrrrr}
            \toprule
            $n$ & 3 & 5 & 7 & 9 & 11 & 15\\
            \midrule
            XAIG & 5 & 11 & 19 & 27 & 34 & 51\\
            AIG & 7 & 17 & 28 & 41 & 52 & 77\\
            \bottomrule
        \end{tabular}
    \end{center}
\end{table}

\subsubsection{MAJ and SORT}
For $\operatorname{MAJ}$ and $\operatorname{SORT}$, our tool improved greatly the best
known circuits following the hybrid approach outlined in~Listing~\ref{lst:hybrid}:
\begin{enumerate}
    \item first, compute $(b_0, \dotsc, b_k)=\operatorname{SUM}(x_1, \dotsc, x_n)$
    (here, $k=\lceil \log_2 (n + 1) \rceil-1$);
    \item then, using a~SAT-based approach, find a~circuit
    that computes the required function out~of $(b_0, \dotsc, b_k)$;
    \item finally, minimize the composition of~the two circuits.
\end{enumerate}
Tables~\ref{table:maj} and~\ref{table:sort} compare the circuits constructed this way
with the best circuits submitted to~the IWLS 2023 Programming Contest.

\begin{table}[tb]
    \caption{Size of~XAIG circuits for $\operatorname{MAJ}_n$: the best circuits submitted
    to~IWLS 2023 Programming Contest and circuits synthesized by~our tool.}
    \label{table:maj}
    \begin{center}
        \begin{tabular}{lrrrrrrr}
            \toprule
            $n$ & 7 & 9 & 11 & 13 & 15\\
            \midrule
            IWLS 2023 &  18 & 29 & 40 & 52 & 68\\
            Our tool, 2024  & 17 & 24 & 31 & 45 & 48\\
            Improvement & 5\% & 17\% & 22\% & 13\% & 29\%\\
            \bottomrule
        \end{tabular}
    \end{center}
\end{table}

\begin{table}[tb]
    \caption{Size of~XAIG circuits for $\operatorname{SORT}_n$: the best circuits submitted
        to~IWLS 2023 Programming Contest and circuits synthesized by~our tool.}
    \label{table:sort}
    \begin{center}
        \begin{tabular}{lrrrrrrr}
            \toprule
            $n$ & 12 & 13 & 14 & 15 & 16\\
            \midrule
            IWLS 2023 & 64 & 90 & 77 & 114 & 109\\
            Our tool, 2024  & 58 & 62 & 68 & 73 & 82\\
            Improvement & 9\% & 31\% & 11\% & 35\% & 24\%\\
            \bottomrule
        \end{tabular}
    \end{center}
\end{table}

\subsubsection{Multipliers}
As~mentioned above, efficient circuits for $\operatorname{SUM}$
allow to~synthesize circuits for various arithmetic functions
as~many of~them use bit summation, one way or~another.
A~prominent example is~$\operatorname{MULT}_n$.
With respect to~circuit size, efficient long integer multiplication
algorithms (like the ones by~\cite{Kar} and
\cite{DBLP:journals/computing/SchonhageS71})
start to~outperform the straightforward grade-school algorithm
only for large values of~$n$. Even when using the grade-school
algorithm with small~$n$, it~is still unclear what is~the best way
of~computing the sum of~the pairwise products of~input bits.
The tool contains a~number of~generators of~efficient multiplier circuits
(which, in~turn, are based on~efficient $\operatorname{SUM}$ circuits described above).

\subsubsection{Arithmetic Functions}
In~the IWLS 2024 Programming Contest, it~is the arithmetic functions
category where we~were able to~achieve the most dramatic improvement
in~circuit size compared to~the best results of~2023 (see the first four
rows of~Table~\ref{table:datasets}). For each of~the arithmetic benchmarks,
we~followed the same two-step approach:
\begin{enumerate}
    \item Convert a~known algorithm for computing
    the corresponding function to~a~circuit.
    When doing this, optimize individual parts
    of~the circuit whenever possible.
    \item Minimize the resulting circuit.
\end{enumerate}

We~illustrate the first step of~this approach for the $\operatorname{DIV}$ function.
Assume that
\[\operatorname{DIV}_n(x_0, \dotsc, x_{n-1}, y_0, \dotsc, y_{n-1})=(z_0, \dotsc, z_{n-1}),\] where $x_0, y_0, z_0$ are the least significant bits
of~the corresponding integers. Let also
\begin{align*}
X&=\operatorname{int}(x_0, \dotsc, x_{n-1}),\\
Y&=\operatorname{int}(y_0, \dotsc, y_{n-1}),\\
Z&=\operatorname{int}(z_0, \dotsc, z_{n-1})
\end{align*}
be~the corresponding integers. Thus,
$0 \le X, Y, Z < 2^n$ and  $Z = \lfloor X/Y \rfloor$.
We~assume also that $Y>0$.
We~follow the grade-school division algorithm: determine the bits of~$Z$
one by~one starting from the most significant bit, each time try subtracting $2^{i}z_iY$
from~$X$. Since $z_i$ is~either~$0$ or~$1$, one does not even need to~multiply by~$z_i$:
$2^iz_iY$ is~either $0$~or~$Y$ shifted by~$i$~positions.
Moreover, all iterations of~the algorithm perform similar checks and one
can precompute the required checks in~advance using the dynamic programming
technique. We~provide the details below.

The algorithm recovers the bits $z_i$ for $i=n-1, \dotsc, 0$.
For the $i$-th iteration, by
\(X' = X - \sum_{j = i + 1}^{n - 1}{2^j z_j Y}\)
we~denote the ``remaining part'' of~$X$ that still needs to~be divided by~$Y$.
Then,
$z_i=1$ if~and only~if $X' \ge 2^i Y$. This~is, in~turn,
equivalent~to
\[Y < 2^{n-i} \land \operatorname{int}(x'_{i}, \dotsc, x'_{n-1}) \ge \operatorname{int}(y_0, \dotsc, y_{n-i-1})\, .\]
Finally, the first~of these two conditions is~
\(\bigvee_{j = n - i}^{n-1}\overline{y_j} .\)

We~check both these conditions as~follows.
To~avoid recomputing $p_i=\bigvee_{j = n - i}^{n-1}\overline{y_j}$ at~every iteration from scratch,
we~precompute $p_{n-1}, \dotsc, p_0$ using the dynamic programming approach:
\(p_{n - 1} = \overline{y_{n - 1}}\)  and \(p_i = p_{i + 1} \lor \overline{y_i}\), for $i=n-2, \dotsc, 0$.

For the second part, we~need to~compare two $(n-i)$-bit integers.
Note that if~$z_i=1$, then one also needs to~update~$X'$.
It~turns out that one can combine the comparison and the subtraction in~the same subcircuit
of~size $5(n-i)$. To~achieve this, we~process the two integers going from the least
significant bit to~the most significant one. For each position, we~apply the Full Adder block to~the three bits: two bits of~the numbers to~be compared and the carry bit from the previous position. This way, the final carry bit~$c_i$ is~equal to~zero if~and only~if
\(\operatorname{int}(x'_{i}, \dotsc, x'_{n-1}) \ge \operatorname{int}(y_0, \dotsc, y_{n-i-1}).\)
Then, $z_i = \overline{c_i} \land \overline{p_i}$.
To~update~$X'$, one uses $n-i$ if-then-else blocks, each having circuit size equal to~$3$.
Overall, the size of~the resulting circuit is~about $4n^2$.

\subsection{Minimizing Existing Circuits}
We~have been able to~synthesize efficient circuits for various
symmetric and arithmetic functions, partly due~to~the fact
that we~knew the exact structure of~these functions. At~the same time,
in~the IWLS 2024 Programming Competition, there were many
benchmarks whose structure was unclear:
for example, 32~benchmarks correspond to~three-output neurons
from the LogicNets project~\cite{DBLP:conf/fpl/UmurogluAFB20}.

In~such cases,
we~were trying to~optimize either a~known circuit or~some
inefficient circuit for the corresponding function.
To~do this, we~combined our SAT-based minimization tool
with other efficient tools like \texttt{ABC}~\cite{DBLP:conf/cav/BraytonM10} and
and \texttt{CIOPS}~\cite{DBLP:conf/aaai/ReichlSS23}: for a~given circuit,
we~applied the three tools repeatedly. This strategy allowed~us
to~further improve about half of~the circuits synthesized for
the neuron benchmarks, see Table~\ref{table:neuron}.

\begin{table}[tb]
    \caption{Size of~XAIG circuits for neuron benchmarks: the best circuits submitted
        to~IWLS 2023 Programming Contest and circuits synthesized by~our tool.}
    \label{table:neuron}
    \begin{center}
        \begin{tabular}{lrrrrrrr}
            \toprule
            benchmark & ex29 & ex58 & ex77 & ex28 & ex54\\
            \midrule
            IWLS 2023 & 644 & 283 & 167 & 398 & 426\\
            Our tool, 2024  & 587 & 280 & 165 & 394 & 421\\
            Improvement & 8\% & 1\% & 1\% & 1\% & 1\%\\
            \bottomrule
        \end{tabular}
    \end{center}
\end{table}

%\section{Conclusion}
%The presented tool has been published online as~an open-source repository and is available at \url{https://github.com/SPbSAT/cirbo}.  We hope that it~will be~useful for the researchers from the SAT community and beyond. Our future plans include extending the functionality of the~tool
%by~adding the support for more features on~circuit satisfiability and verification.

\section*{Acknowledgments}
Research of~the authors 1, 2, 3, 5, 7, 10, and 11 is supported by Huawei (grant TC20211214628).
Research of~the author~2 is~additionally supported by~the grant 075-15-2022-289 for creation and development of Euler International Mathematical Institute, and by the Foundation for National Technology Initiative's Projects Support (agreement 70-2021-00187).

\bibliography{references}

\begin{thebibliography}{18}
\providecommand{\natexlab}[1]{#1}

\bibitem[{Biere, Heljanko, and Wieringa(2011)}]{BiereHeljankoWieringa2011}
Biere, A.; Heljanko, K.; and Wieringa, S. 2011.
\newblock AIGER 1.9 And Beyond.
\newblock Technical report, FMV Reports Series, JKU Linz, Austria.

\bibitem[{Biere et~al.(2021)Biere, Heule, van Maaren, and
  Walsh}]{DBLP:series/faia/336}
Biere, A.; Heule, M.; van Maaren, H.; and Walsh, T., eds. 2021.
\newblock \emph{Handbook of Satisfiability - Second Edition}, volume 336 of
  \emph{Frontiers in Artificial Intelligence and Applications}.
\newblock {IOS} Press.

\bibitem[{Brayton and Mishchenko(2010)}]{DBLP:conf/cav/BraytonM10}
Brayton, R.~K.; and Mishchenko, A. 2010.
\newblock {ABC:} An Academic Industrial-Strength Verification Tool.
\newblock In \emph{{CAV}}, volume 6174 of \emph{Lecture Notes in Computer
  Science}, 24--40. Springer.

\bibitem[{Demenkov et~al.(2010)Demenkov, Kojevnikov, Kulikov, and
  Yaroslavtsev}]{DBLP:journals/ipl/DemenkovKKY10}
Demenkov, E.; Kojevnikov, A.; Kulikov, A.~S.; and Yaroslavtsev, G. 2010.
\newblock New upper bounds on the Boolean circuit complexity of symmetric
  functions.
\newblock \emph{Inf. Process. Lett.}, 110(7): 264--267.

\bibitem[{Find et~al.(2016)Find, Golovnev, Hirsch, and
  Kulikov}]{DBLP:conf/focs/FindGHK16}
Find, M.~G.; Golovnev, A.; Hirsch, E.~A.; and Kulikov, A.~S. 2016.
\newblock A Better-Than-3n Lower Bound for the Circuit Complexity of an
  Explicit Function.
\newblock In \emph{{FOCS}}, 89--98. {IEEE} Computer Society.

\bibitem[{Find et~al.(2023)Find, Golovnev, Hirsch, and
  Kulikov}]{DBLP:journals/cc/FindGHK23}
Find, M.~G.; Golovnev, A.; Hirsch, E.~A.; and Kulikov, A.~S. 2023.
\newblock Improving 3N Circuit Complexity Lower Bounds.
\newblock \emph{Comput. Complex.}, 32(2): 13.

\bibitem[{Haaswijk et~al.(2017)Haaswijk, Testa, Soeken, and
  Micheli}]{DBLP:conf/ismvl/HaaswijkTSM17}
Haaswijk, W.; Testa, E.; Soeken, M.; and Micheli, G.~D. 2017.
\newblock Classifying Functions with Exact Synthesis.
\newblock In \emph{{ISMVL}}, 272--277. {IEEE} Computer Society.

\bibitem[{Ignatiev, Morgado, and
  Marques{-}Silva(2018)}]{DBLP:conf/sat/IgnatievMM18}
Ignatiev, A.; Morgado, A.; and Marques{-}Silva, J. 2018.
\newblock PySAT: {A} Python Toolkit for Prototyping with {SAT} Oracles.
\newblock In \emph{{SAT}}, volume 10929 of \emph{Lecture Notes in Computer
  Science}, 428--437. Springer.

\bibitem[{Karatsuba and Ofman(1963)}]{Kar}
Karatsuba, A.~A.; and Ofman, Y. 1963.
\newblock Multiplication of many-digital numbers by automatic computers.
\newblock \emph{Dokl. Akad. Nauk SSSR}, 145(2): 293--294.

\bibitem[{Kojevnikov, Kulikov, and
  Yaroslavtsev(2009)}]{DBLP:conf/sat/KojevnikovKY09}
Kojevnikov, A.; Kulikov, A.~S.; and Yaroslavtsev, G. 2009.
\newblock Finding Efficient Circuits Using SAT-Solvers.
\newblock In \emph{{SAT}}, volume 5584 of \emph{Lecture Notes in Computer
  Science}, 32--44. Springer.

\bibitem[{Kulikov, Pechenev, and Slezkin(2022)}]{DBLP:conf/mfcs/KulikovPS22}
Kulikov, A.~S.; Pechenev, D.; and Slezkin, N. 2022.
\newblock SAT-Based Circuit Local Improvement.
\newblock In \emph{{MFCS}}, volume 241 of \emph{LIPIcs}, 67:1--67:15. Schloss
  Dagstuhl - Leibniz-Zentrum f{\"{u}}r Informatik.

\bibitem[{Li and Yang(2022)}]{DBLP:conf/stoc/Li022}
Li, J.; and Yang, T. 2022.
\newblock $3.1n-o(n)$ circuit lower bounds for explicit functions.
\newblock In \emph{{STOC}}, 1180--1193. {ACM}.

\bibitem[{Reichl, Slivovsky, and Szeider(2023)}]{DBLP:conf/aaai/ReichlSS23}
Reichl, F.; Slivovsky, F.; and Szeider, S. 2023.
\newblock Circuit Minimization with QBF-Based Exact Synthesis.
\newblock In \emph{{AAAI}}, 4087--4094. {AAAI} Press.

\bibitem[{Santhanam(2022)}]{DBLP:conf/fsttcs/Santhanam22}
Santhanam, R. 2022.
\newblock Why {MCSP} Is a More Important Problem Than {SAT} (Invited Talk).
\newblock In \emph{{FSTTCS}}, volume 250 of \emph{LIPIcs}, 2:1--2:1. Schloss
  Dagstuhl - Leibniz-Zentrum f{\"{u}}r Informatik.

\bibitem[{Sch{\"{o}}nhage and
  Strassen(1971)}]{DBLP:journals/computing/SchonhageS71}
Sch{\"{o}}nhage, A.; and Strassen, V. 1971.
\newblock Schnelle Multiplikation gro{\ss}er Zahlen.
\newblock \emph{Computing}, 7(3-4): 281--292.

\bibitem[{Soeken et~al.(2018)Soeken, Riener, Haaswijk, and
  Micheli}]{DBLP:journals/corr/abs-1805-05121}
Soeken, M.; Riener, H.; Haaswijk, W.; and Micheli, G.~D. 2018.
\newblock The {EPFL} Logic Synthesis Libraries.
\newblock \emph{CoRR}, abs/1805.05121.

\bibitem[{Tseitin(1968)}]{Tseitin70}
Tseitin, G. 1968.
\newblock On the complexity of derivation in propositional calculus.
\newblock \emph{Studies in Constructive Mathematics and Mathematical Logic},
  115–--125.

\bibitem[{Umuroglu et~al.(2020)Umuroglu, Akhauri, Fraser, and
  Blott}]{DBLP:conf/fpl/UmurogluAFB20}
Umuroglu, Y.; Akhauri, Y.; Fraser, N.~J.; and Blott, M. 2020.
\newblock LogicNets: Co-Designed Neural Networks and Circuits for
  Extreme-Throughput Applications.
\newblock In \emph{{FPL}}, 291--297. {IEEE}.

\end{thebibliography}

\end{document}